\documentclass[usenatbib]{mn2e}
\usepackage{psfig}

\newcommand{\spose}[1]{\hbox to 0pt{#1\hss}}
\newcommand{\approxpropto}{\mathrel{\spose{\lower 3pt\hbox{$\sim$}}
	\raise 2.0pt\hbox{$\propto$}}}
\def\approxgt{\mathrel{\spose{\lower 3pt\hbox{$\sim$}}
	\raise 2.0pt\hbox{$>$}}}
\def\approxlt{\mathrel{\spose{\lower 3pt\hbox{$\sim$}}
	\raise 2.0pt\hbox{$<$}}}

\newcommand{\etal}{et al.{}}
\newcommand{\keV}{\mbox{\rm keV}}

\newcommand{\Msun}{\mbox{$\rm M_{\odot}$}}

\newcommand{\Nbody}{\mbox{$N$-body}}

\title[The power spectrum amplitude from clusters]
{The power spectrum amplitude from clusters revisited: \\ 
$\sigma_8$ using simulations with preheating and cooling}

\author[P.~T.~P.~Viana \etal]{Pedro T. P. Viana$^{1,2}$, Scott T.~Kay,$^3$ 
Andrew R.~Liddle,$^3$ Orrarujee Muanwong,$^{3,4}$ \cr and Peter
A.~Thomas$^3$ \\
\vspace*{-6pt} {\small \em $^1$Centro de Astrof\'{\i}sica, 
Universidade do Porto, Rua das Estrelas, 4150-762 Porto, Portugal}\\
\vspace*{-6pt} {\small \em $^2$Departamento de Matem\'{a}tica Aplicada, 
Faculdade de Ci\^{e}ncias, 
Universidade do Porto, Rua do Campo Alegre, 687,}\\
{\small \em  ~~~~4169-007 Porto, Portugal}\vspace*{-6pt}\\
{\small \em $^3$Astronomy Centre, University of Sussex, Falmer, 
Brighton, BN1\,9QJ}\vspace*{-6pt}\\
{\small \em $^4$Department of Physics, Faculty of Science, Khon Kaen
University, Khon Kaen, 40002 Thailand}}

\date{\today}

\begin{document}
\journal{Preprint astro-ph/0211090} 
 
\maketitle

\begin{abstract}
The amplitude of density perturbations, for the currently-favoured
$\Lambda$CDM cosmology, is constrained using the observed properties 
of galaxy clusters. The catalogue used is that of Ikebe et al.~(2002).  
The cluster temperature to mass relation is obtained via 
$N$-body/hydrodynamical simulations including radiative cooling 
and preheating of cluster gas, which we have previously shown
to reproduce well the observed temperature--mass relation in the
innermost parts of clusters (Thomas et al.~2002). We generate and 
compare mock catalogues via a Monte Carlo method, which allows us 
to constrain the relation between X-ray temperature and luminosity, 
including its scatter, simultaneously with cosmological parameters.
We find a luminosity--temperature relation in good agreement with the 
results of Ikebe et al.~(2002), while for the matter power spectrum
normalization, we find $\sigma_8 = 0.78_{-0.06}^{+0.30}$ at 95 per
cent confidence for $\Omega_0 = 0.35$. Scaling to WMAP's central value
of $\Omega_0 = 0.27$ would give a best-fit value of $\sigma_8 \simeq
0.9$.
\end{abstract}

\begin{keywords}
methods: \Nbody\  simulations -- hydrodynamics -- X-rays: galaxies:
clusters -- galaxies: clusters: general
\end{keywords}

\section{Introduction}

It has recently become apparent that traditional hydrodynamical
simulations, where the gas is only allowed to heat adiabatically and
through shocks, have difficulties in matching observations in the
central regions of clusters, with a significant underestimation of the
temperature corresponding to a given cluster mass. This is potentially
important for attempts to use the observed temperature function of
clusters to constrain the matter power spectrum on short scales, a
topic which has been studied by many authors over the years
(\citealt{E89}; \citealt{HA91}; \citealt{OB92}; \citealt*{WEF93};
\citealt*{EKF96}; \citealt{VL96}; \citealt[][hereafter VL99]{VL99};
\citealt{H97}, \nocite{H00}2000; \citealt{BSB00}; \citealt*{PSW01};
\citealt{W01}), most of which use hydrodynamical simulations to relate
mass to temperature. Such concerns have been given further impetus by
a recent paper by \citet{S01}, who used an observed relationship
between cluster temperature and mass \citep*{FRB01}, rather than one
derived from hydrodynamical simulations, to find a normalization for
the matter power spectrum significantly lower than that of earlier
works.  In a recent paper (\citealt{Tetal02}) we showed that the
inclusion of extra gas physics, namely radiative cooling of the gas
and possible preheating of the gas before cluster formation, can bring
simulations into good agreement with recent {\it Chandra} observations
of the cores of clusters (\citealt{ASF01}), suggesting that these may
be crucial ingredients in obtaining an accurate description of
clusters.

In this paper, we derive a constraint on the matter power spectrum 
normalization $\sigma_8$ in a way which improves on previous work in several 
ways. On the theoretical side, we incorporate the temperature--mass
relationship, and its scatter, as obtained from the simulations described above. 
On the observational side, we compare with the data published in Ikebe et 
al.~(2002), whose raw catalogue contains around one hundred clusters, most with 
data from both {\it ROSAT} and {\it ASCA}. Finally, on the data analysis side we 
use a novel approach, whereby Monte Carlo simulations are used to generate 
mock galaxy cluster catalogues, which through comparison with the 
data published in Ikebe et al.~(2002) lead to a simultaneous 
constraint on the X-ray temperature to luminosity relationship, 
including its scatter, and on the matter power spectrum normalization 
$\sigma_8$. 

\section{The observed cluster catalogue}

The galaxy cluster catalogue containing the best available 
X-ray data is that compiled by Ikebe et al.~(2002) and 
Reiprich \& B\"ohringer (2002). The master catalogue contains 
106 clusters, selected by their X-ray {\it ROSAT} 
flux from available cluster catalogues, with 88 among them 
having been observed by {\it ASCA}. Imposing a flux cut in 
the {\it ROSAT} [0.1, 2.4] keV band of 
$2.0\times10^{-11}\;{\rm erg}\,{\rm s}^{-1}\,{\rm cm}^{-2}$, 
a flux-limited sample of 63 clusters is then obtained, 
called {\it HIFLUGCS}, which is claimed to be statistically complete 
(Reiprich \& B\"ohringer 2002). Ikebe et al.~(2002) use a 
slightly different sample in their analysis, obtained by 
excluding the two lowest temperature clusters from 
{\it HIFLUGCS}, ending up with a sample of 61 
clusters with X-ray temperatures ranging from 1.4 keV up 
to 11 keV. Among these, 56 have X-ray temperatures 
derived by Ikebe et al.~(2002) from {\it ASCA} data by 
means of a two-temperature model that takes into account 
a possible contribution from a cooler component at the 
cluster core. 

In order to define the observed cluster sample with which 
to compare the artificially-generated cluster catalogues, we 
will impose more restrictive selection criteria on {\it HIFLUGCS} 
than Ikebe et al.~(2002) did. We will only consider clusters 
with measured X-ray flux in the [0.1, 2.4] keV band above 
$2.2\times10^{-11}\;{\rm erg}\,{\rm s}^{-1}\,{\rm cm}^{-2}$, 
X-ray temperature higher than 2 keV, and a redshift between 
0.03 and 0.10 (when performing tests, we found that 
including clusters with $z<0.03$ seems to lead to an 
increase in the best-fit $\sigma_{8}$ by a few per cent). 
These ranges were chosen to maximize confidence 
in completeness of the sample, minimize cosmic variance, 
and because the luminosity--temperature relation is expected 
to deviate from a power law below 2 keV due to 
non-gravitational physics.

So that we can account for the measurement errors both 
in flux and temperature, which can lead to incompleteness 
effects when imposing either flux or temperature criteria 
in the sample selection procedure, we used Monte Carlo 
simulations to generate 40 realizations of the {\it HIFLUGCS} 
catalogue, with the measurement errors in flux and temperature 
modelled as Gaussian distributed. We then imposed our cluster 
selection criteria, described above, on these catalogues to 
obtain a set of 40 observed cluster samples, with very similar 
but not identical numbers of clusters, representing different 
possible realizations of the chosen observed cluster sample. 

We performed extensive tests to determine the minimum number 
of Monte Carlo realizations of {\it HIFLUGCS} that should be 
generated, so as to properly account for the effect of the 
measurement errors on the distribution of the cluster 
properties, within the observed data sample which we will use 
to compare with the artificially-generated cluster catalogues. 
We found that 40 realizations are enough, and increasing 
their number to 200 or 1000 has a negligible effect 
both on the typical distribution of cluster properties 
and on the final probability distribution for $\sigma_{8}$. 
We also generated bootstrap realizations of {\it HIFLUGCS}, to 
determine whether the flux and temperature measurement errors 
provided by Ikebe et al.~(2002) were realistic. They seem to 
be, given that the bootstrap realizations share the same mean 
statistical properties as the Monte Carlo ones, leading to 
negligible differences in the final probability distribution 
of $\sigma_{8}$ values. Finally, there does not seem to be any 
systematic shift in the mean statistical properties of both 
the Monte Carlo and Bootstrap catalogues with relation to 
{\it HIFLUGCS}, which is reflected by the fact that our 
result on $\sigma_{8}$ does not change even if we 
just apply our selection criteria to {\it HIFLUGCS}, and then  
compare the resulting cluster sample with the 
artificially-generated cluster catalogues.

\section{The mock cluster catalogues}

The direct simulation of X-ray cluster catalogues from
hydrodynamical simulations is beyond present computational means due to the 
excessive number of particles required to 
obtain statistically-robust cluster abundances with temperatures 
above a few keV. Instead, we appeal to the method used by 
Holder, Haiman \& Mohr (2001), which is to use generalized mass 
functions of dark matter haloes to generate catalogues of 
clusters identified by their redshift and mass, and then estimate 
their X-ray temperatures using the mass--temperature relation of 
clusters in hydrodynamical simulations. With relation to previous 
work, the main improvement in this paper is the use of a 
mass--temperature relation that is drawn from simulations 
with more detailed models of the intracluster gas physics than have 
previously been implemented, and which closely match the X-ray properties 
of observed clusters (Thomas et al.~2002; Muanwong et al.~2002, hereafter 
MTKP02). 

\subsection{The mass function}

Currently no standard definition of a dark matter halo exists, although 
it is convenient to define a halo as an overdense concentration of matter 
using the results of the spherical top-hat collapse model 
(STHCM; e.g.~\citealt{Pee93}; although see also \citealt{SMT01}).
For $\Omega=1$, the boundary of a halo predicted by the STHCM contains 
a mean internal overdensity of 18$\pi^2 \approx 178$ relative to the critical 
density. This result has led many authors to define haloes using an 
overdensity contrast of 200 (which we take as our fiducial case). Note
that even with the current generation of X-ray satellites it is not feasible to 
measure spatially-resolved properties of clusters to such large radii.

A comprehensive study of the mass function of cold dark matter haloes
was carried out by Jenkins et al.~(2001, hereafter Jen01), who compared 
results from the largest
$N$-body simulations available (the Hubble Volume simulations
simulated by the Virgo Consortium, which used sufficiently large
volumes to obtain reliable abundances of haloes on scales
corresponding to rich clusters of galaxies) to the mass function 
predicted by \citet{PS74}. They demonstrated that
the simulated mass function predicts a lower abundance of haloes at
low masses than the Press--Schechter function, but a higher abundance
at high masses.  Although they did not investigate the cause of this
discrepancy, they pointed out that
the Press--Schechter {\it ansatz} that all mass is contained
in bound objects is untrue in the simulations for conventional halo
definitions.

Jen01 produced fits to simulated mass functions using two different
estimators: the spherical-overdensity (SO) and friends-of-friends
(FOF) algorithms. The first case, as implemented by \citet{LC94},
finds and ranks the densest dark matter particles and, starting from
the densest, grows a sphere until the mean internal density equals
some multiple of the critical density, $\rho_{\rm cr}$, $\left< \rho
\right> = \Delta \rho_{\rm cr}$.  Particles within this halo are then
removed from the list and the procedure is repeated until all haloes
are found down to a given mass limit. The FOF algorithm \citep{DEFW85}
links particles together using a fixed linking length of $bn^{-1/3}$,
where $n$ is the mean particle density.  FOF does not impose
spherical symmetry on the shapes of the haloes (which are typically triaxial)
but can sometimes link together haloes which are in close proximity.
It is important to use a consistent definition for
cluster masses to define both the mass function and the
mass--temperature relation: failure to do so can lead to errors of
10 per cent in the derived value of $\sigma_8$.

A further result from the Jen01 analysis was that the mass functions,
when expressed as a function of $\ln(\sigma^{-1})$ (where $\sigma(M)$
is the generalization of $\sigma_8$ to any mass-scale), are
independent of cosmology if haloes are defined using either a fixed 
linking length (e.g. $b=0.2$) in the FOF case or defining the 
spherical-overdensity threshold with respect to the mean background 
density (e.g. $\Delta=180\Omega_0$) in the SO case. This was 
confirmed by \citet[][hereafter Evr02]{Evr02}, who also 
provided fits (as a function of $\Omega$) to simulated mass 
functions using a SO algorithm with $\Delta=200$ 
(i.e. overdensity measured with respect to the 
critical density). For this paper, we adopt $M_{200}$ 
as the fiducial definition of cluster mass, and use the Evr02 fits 
to estimate the mass function at different $\Omega(z)$. We
have checked that our method for measuring cluster masses from the
simulations (required for the calibration of the mass--temperature
relation) produces almost identical results to the SO method used by
Jen01 and Evr02 (the median difference in halo masses is less than 0.5
per cent).

\subsection{The mass--temperature relation}

In this section, we use results drawn from simulations carried out
using the {\sc hydra}\footnote{http://hydra.susx.ac.uk/}
\Nbody/hydrodynamics code (\citealt*{CTP95}; \citealt{PC97}) on the
Cray T3E computer at the Edinburgh Parallel Computing Centre as part
of the Virgo Consortium\footnote{http://virgo.susx.ac.uk/} programme
of investigations into structure formation in the Universe.  Details
of the method and choice of simulation parameters were discussed by
MTKP02; we summarize details pertinent to
the results of this paper below.

We adopt the currently-favoured $\Lambda$CDM cosmological model,
setting the density parameter $\Omega_0=0.35$, cosmological constant
$\Omega_{\Lambda}=0.65$, baryon density $\Omega_{\rm b}=0.038$, Hubble
parameter $h=0.71$ and linear power spectrum shape parameter
$\Gamma=0.21$.  The purpose of this paper is to provide constraints on
$\sigma_8$ and so it may seem premature to pick one particular value
for our simulations.  However, the mass--temperature relation of
clusters is largely independent of $\sigma_8$.  The simulations
presented in MTKP02 use $\sigma_8=0.9$; we have subsequently repeated
one of the simulations with a lower normalization, $\sigma_8=0.7$,
and find an identical relation within the uncertainties.

MTKP02 presented 3 simulations which differed in the way in which
the gas was heated and cooled. In the first simulation, a {\it
Non-Radiative} model, the gas could undergo heating by adiabatic
compression and shocks but could not cool radiatively.  Consequently,
the resulting clusters are far too luminous for their mass and so do
not agree with observed X-ray scaling relations (MTKP02). We do not use
results from this simulation. 

In the {\it Radiative} simulation, gas was able to cool radiatively
using the collisional ionization equilibrium tables of \citet{SD93}.
Cooled material was permitted to form stars, removing low-entropy
material with short cooling times from the centres of the clusters.
Finally, in the {\it Preheating} simulation (which also includes cooling), the 
specific thermal energy of the gas was raised by 1.5 keV per 
particle at $z=4$, to crudely model the effects
of energy injection by galactic winds.  Both models reproduce key X-ray
cluster scaling relations at $z=0$, although the former predicts too
much cooled gas (i.e.~stars and galaxies) compared to observations and the 
latter too little.

We estimate the X-ray temperature of each cluster by weighting 
the contribution from each hot gas ($T>10^{5}$K) particle 
by its bolometric flux
\begin{equation}
T_X = \frac{\Sigma_i m_i\rho_i\Lambda_{\rm bol}(Z,T_i)T_i}
          {\Sigma_i m_i\rho_i\Lambda_{\rm bol}(Z,T_i)}.
\label{eq:ktx}
\end{equation}
Here, $m_i$, $\rho_i$ and $T_i$ are the mass, density and temperature
of the particles, $Z=0.3Z_{\odot}$ is their metallicity and 
$\Lambda_{\rm bol}$ is the bolometric cooling function from 
\citet{SD93}. Adopting a soft-band cooling function (appropriate for 
{\it ROSAT} observations) makes no significant difference to the 
estimated temperature. Many clusters show enhanced emission from
the cluster core that has a lower temperature than the cluster mean (MTKP02).
For this reason, we present results for the mass--temperature relation
both including and excluding the X-ray emission from within the
`cooling radius', defined as the radius within which the mean cooling time of 
the gas is 6\,Gyr. The latter results are referred to as `cooling-flow 
corrected'.

\begin{table}
\caption{Power-law fits to the simulated mass--temperature relations of
X-ray clusters: cluster sample; number of clusters in sample; slope of
relation, $s$; rms dispersion in temperature about best-fit (see
text); value of $M_{200}/10^{14}h^{-1}\Msun$ at 3\,keV; value of
$M_{200}/10^{14}h^{-1}\Msun$ at 6\,keV.}
\label{tab:mtfit}
\begin{tabular}{lccccr}
\hline
Sample& $N$& $s$& rms& $M_{200}@3$& $M_{200}@6$\\
\hline
\multicolumn{6}{c}{All data}\\
\hline
{\it Radiative}                 & 36& 1.80& 0.092& 2.9& 10.1\\
{\it Preheating}, $\sigma_8=0.9$& 31& 1.59& 0.056& 2.4& 7.3\\
{\it Preheating}, $\sigma_8=0.7$& 12& 1.75& 0.049& 2.1& 7.1\\
{\it Preheating}, $\sigma_8=$any& 43& 1.61& 0.053& 2.4& 7.3\\
\hline
\multicolumn{6}{c}{Cooling-flow corrected}\\
\hline
{\it Radiative}                 & 36& 1.55& 0.079& 2.3& 6.8\\
{\it Preheating}, $\sigma_8=0.9$& 31& 1.51& 0.054& 2.2& 6.2\\
{\it Preheating}, $\sigma_8=0.7$& 12& 1.70& 0.040& 2.4& 7.7\\
{\it Preheating}, $\sigma_8=$any& 43& 1.54& 0.049& 2.2& 6.4\\
\hline
\end{tabular}
\end{table}

In Table~\ref{tab:mtfit}, we list parameters for the straight-line 
relation of the form
\begin{equation}
\log(kT/\keV)={\rm const}+(1/s)\;\log(M_{200}/h^{-1}\Msun)
\end{equation}
that minimizes the dispersion in temperature for all clusters with
$\log(M_{200}/h^{-1}\Msun)>14$. The column labelled ``rms'' gives the 
root-mean-square dispersion in the log of 
temperature (for $N-2$ degrees of freedom) about the best-fit line.
We have also measured this dispersion for clusters in a lower mass
range, $13.7<\log(M_{200}/h^{-1}\Msun)<14$, and find very similar 
values. Hence we will assume in our analysis that the dispersion is 
independent of mass.

The final two columns of Table~1, labelled $M_{200}@3$ and $M_{200}@6$, give 
the values of the mass, in units of $10^{14}h^{-1}\;\Msun$, for the best-fit
relation at temperatures of 3 and 6\,keV.  The numbers in the $M_{200}@3$
column are mostly very similar to each other, except for the top entry
for clusters in the {\it Radiative} simulation without the
cooling-flow correction.  The presence of cool gas in the cores of
these clusters lowers the emission-weighted temperature and hence
raises $M_{200}@3$.  The slope of the temperature--mass relation for
$\sigma_8=0.7$ is higher than that for $\sigma_8=0.9$ but the two are
in agreement to within the errors; with only 12 clusters covering a
limited mass-range, the formal 1-sigma error in the slope for the
$\sigma_8=0.7$ clusters is about $\pm0.4$. 
The predictions for the normalizations of the relations at 6\,keV are
less certain, especially for $\sigma_8=0.7$, because they require a
degree of extrapolation beyond the temperature range of the simulated data.
For this reason, the difference between the cooling-flow corrected
normalizations at 6\,keV for $\sigma_8=0.7$ and $\sigma_8=0.9$ should not be
taken too seriously.  We use the combined catalogue for our analysis
in the next section, but note that very similar results are obtained
if we use the $\sigma_8=0.9$ relation instead.

In Table~\ref{tab:mtsim} we present results from several earlier studies of
the mass--temperature relation in non-radiative simulations. Note that these 
results have been obtained by rescaling, when needed, the cluster mass to 
$M_{200}$ (using a NFW profile: Navarro, Frenk \& White 1995, 1996, 1997) 
and to the cosmology being considered here (as in BN98). Clearly 
there is a wide range of normalizations. This mainly results from the 
different resolutions of the simulations (though in the case of EMN96 their 
method of temperature estimation also plays a part). Also, on average, at 
fixed temperature the cluster masses in Table~\ref{tab:mtsim} are higher 
than those in Table~1. This is due to the absence of radiative cooling; 
cluster cores are full of dense, cold gas with short 
cooling times and this leads to low emission-weighted temperatures. 
This problem is largely overcome in the {\it Radiative} and {\it Preheating}
simulations and can be reduced even further by the omission of the
cooling-flow component. For comparison, Viana \& Liddle (1996) and VL99 used a 
normalization for the present-day mass--temperature relation which corresponds 
to $M_{200}=10.1\times10^{14}h^{-1}\Msun(kT/6\keV)^{1.5}$, based on a
simulation of a single high-mass cluster from White et al.~(1993b). This
agrees well with the results of BN98 and ME01, but lies well above the 
values found by EMN96 and in the {\it Radiative} and {\it Preheating}
simulations reported in this paper. This change in normalization forces the 
estimate of $\sigma_8$ downwards.

\begin{table}
\caption{Mass-temperature relations of X-ray clusters from previous
simulations: paper (\citealt*[EMN96]{EMN96}; \citealt[BN98]{BN98};
\citealt[T01]{T01}; \citealt[ME01]{ME01}; 
slope of relation, $s$; value of $M_{200}/10^{14}h^{-1}\Msun$ at 3\,keV; value 
of
$M_{200}/10^{14}h^{-1}\Msun$ at 6\,keV.}
\label{tab:mtsim}
\begin{tabular}{llccr}
\hline
Paper&& $s$& $M_{200}@3$& $M_{200}@6$\\
\hline
EMN96& soft band&  1.50& 2.3& 6.5\\
BN98&  bolometric& 1.50& 3.6& 10.2\\ 
T01&   bolometric& 1.50& 2.5& 7.1\\
ME01&  bolometric& 1.39& 4.0& 10.6\\
\hline
\end{tabular}
\end{table}

\subsection{Mock catalogue construction}

We are now in a position to be able to combine the 
Evr02 fits to the mass function with the information on the 
cluster mass--temperature relation from the hydrodynamical 
simulations, to produce mock cluster catalogues with information 
on cluster redshift, mass and X-ray temperature. 

We take the present-day shape of the matter power spectrum to 
be well approximated by that of a cold dark matter model, with 
scale-invariant primordial density perturbations and effective 
shape parameter, $\Gamma=0.18$. This is the favoured value of
$\Gamma$ from a joint analysis of the 2dF (Percival et al.~2001) 
and SDSS (Szalay et al.~2001; Dodelson et al.~2002) data, 
when accounting for both statistical and systematic 
uncertainties (the allowed interval for $\Gamma$ is 
[0.08, 0.28] and we confirmed that varying $\Gamma$ within this 
interval does not significantly change the final results; changing $\Gamma$ to 
either 0.08 or 0.28 leads to a variation of only 0.02 in the best-fit 
$\sigma_{8}$, with a higher $\Gamma$ implying a higher $\sigma_{8}$.). 

We begin by estimating the mean number of clusters as a function
of mass ($M_{200}$) and redshift, using the Evr02 fits to the 
mass function, for each value of $\sigma_{8}$ over the interval 
of interest. Our redshift bins cover the
range [0.03,0.10] in intervals of 0.001, and our mass bins cover
the range $[0.1,\,2.0]\times10^{15}\,h^{-1}\;\Msun$ in 
logarithmically-spaced intervals of 0.01. (We have checked 
that our results are insensitive to smaller bin-sizes.)
The initial mock cluster catalogues are then produced by attributing  
to each ($z$, $M_{200}$) bin, a number of clusters drawn from a 
Poisson distribution whose mean is that predicted by the Evr02 
fits to the mass function. We assign a mass and redshift to each individual
cluster by randomly drawing the two quantities from a
quadratic distribution that best reproduces the variation in the 
cluster numbers in the neighbourhood of that bin. In this manner,
we produce 1000 mock catalogues for each interesting value of $\sigma_{8}$. 
Through extensive tests we found that such number is  
enough to properly account for the effect of the Poisson noise, as 
increasing the number of mock catalogues per $\sigma_{8}$ to 
e.g. 10000 had a negligible effect on the final probability 
distribution for $\sigma_{8}$. 

Each cluster in the catalogues is given an X-ray temperature, randomly 
drawn from a Gaussian distribution in ($\log_{10}M_{200}$,$\,\log_{10}kT$), 
with mean obtained by substituting the cluster mass [multiplied by 
$H(z)/H_{0} \propto\sqrt{\Omega_{0}(1+z)^{3}+(1-\Omega_{0})}$ 
to account for the redshift evolution in the normalization of 
the cluster X-ray temperature to mass relation: see Mathiesen \& Evrard 2001] 
in expression (2), while the dispersion is assumed to be independent of mass. 
We fix the present-day normalization, slope and dispersion of the 
mass--temperature relation using the
joint cluster catalogue obtained from the {\it Preheating} simulations, 
where the X-ray temperatures were cooling-flow corrected. (Using the parameters 
deduced from the {\it Radiative} simulations does not change the final results 
significantly.) Our method approximately reproduces that used by Ikebe et 
al.~(2002) to estimate the observed cluster temperatures. 
We then exclude from the 1000 mock catalogues any cluster whose X-ray 
temperature does not exceed 2 keV. 

To compare our simulated catalogues with the data we still need to
impose the chosen flux selection criterion, which forces us to
use a relation between X-ray luminosity (in the [0.1, 2.4] keV 
rest-frame band) and temperature. In order to be consistent, we determine
this relation from the data simultaneously with $\sigma_8$ (see also Diego et 
al.~2001). We take 
it to be a power-law of the form
\begin{equation}
\log_{10}(L_{\rm X}/h^{-2}\;{\rm erg}\,{\rm s}^{-1})=A+\alpha\log_{10}(kT/{\rm 
keV})\,,
\end{equation}
with a dispersion $\sigma_{\log_{10}L_{\rm X}}$ taken to be independent of 
temperature, and construct a grid of values (with dimensions $21 \times 31  
\times 16$) of the normalization $A$, slope 
$\alpha$ and dispersion. For each point in this grid, and for every one 
of the 1000 catalogues available for each $\sigma_8$, we create 
50 realizations of the luminosity (extensive tests have shown that 
such number is enough to lead to a dense coverage of the range of 
possible luminosity distributions, and increasing 
the number of realizations to e.g. 200 had a negligible effect on the 
final probability distribution for $\sigma_{8}$) 
for every cluster by randomly drawing from a Gaussian distribution 
in ($\log_{10}kT$, $\log_{10}L_{\rm X}$) with the appropriate mean and 
dispersion. Every cluster then has its X-ray flux in the rest-frame 
[0.1, 2.4] keV band derived, from which the flux in 
the observed [0.1, 2.4] keV band is estimated using K-correction formulae. The 
flux 
limit of $2.2 \times10^{-11}\;{\rm erg}\,{\rm s}^{-1}\,{\rm cm}^{-2}$ 
is then imposed. This generates a set of 50000 mock catalogues for each 
combination of the four parameters we wish to estimate from the data. In all, 
over twenty five billion mock catalogues were generated.

\section{Results}

We are now in possession of an ensemble of catalogues representing the observed 
data set, and a collection of mock catalogues for different values of both 
$\sigma_8$ and the parameters that characterize the X-ray 
luminosity--temperature relation. We chose to perform the comparison between the 
observed and theoretical catalogues via a (three-way) 2D Kolmogorov--Smirnov 
test. This test is a generalization to two-dimensional distributions of the 
traditional Kolmogorov--Smirnov test, and is due to 
Fasano \& Franceschini~(1987), following an earlier idea of 
Peacock~(1983). A very good description of the 2D Kolmogorov--Smirnov test 
can be found in Press et al.~(1992). In order to calculate the probability 
of each set of 4 free parameters being the correct one, we compare each of 
the observed catalogues with each mock catalogue,  
and then add the probabilities of each pair of catalogues 
being drawn from the same underlying distribution of cluster 
properties. The probability is taken to be zero if the two catalogues 
being compared do not have the same number of clusters, 
otherwise it is given by the product of the probabilities that result 
from applying the 2D Kolmogorov--Smirnov test to the 3 available 
distributions of cluster properties --- ($z,kT$), ($z,L_{\rm X}$) 
and ($kT,L_{\rm X}$). The set of free parameters considered most 
correct will thus be the one that most often closely reproduces the 
observed distribution of the cluster properties ($z,kT,L_{\rm X}$). 

\subsection{Methodology tests}

As far as we are aware the 2D Kolmogorov--Smirnov test has not previously been
applied in the same context as here.  We chose to employ the KS test rather than 
the widely-used likelihood method because it allows a fairly simple 
incorporation of the selection effects entering the observations, and can allow 
for scatter in the cluster relations. Ikebe et al.~(2002) employed the 
likelihood function, but did not include scatter in the $M$--$T$ relation, 
though Pierpaoli et al.~(2003) were able to include the scatter in a likelihood 
analysis. We favour the 2D KS test because of its ease of implementation, though 
we do not expect it to lead to significantly different results from the 
likelihood method.

It is clearly important to compare the 2D KS test to
the likelihood method. We do this by
applying the two methods to a simplified situation using 1000 mock observational 
cluster catalogues (we
checked that generating more does not affect the results of the comparison)
produced using the Evr02 fits to the mass function in the same manner as
described in subsection 3.3, with each cluster being characterized by its
redshift, $z$, and mass, $M_{200}$. The assumed fiducial model had  
$\Omega_{0}=0.35$, $\sigma_{8}=0.8$ and $\Gamma=0.18$.  The
sky coverage was the same as that of {\it HIFLUGCS} and the redshift interval
considered was $0.03<z<0.10$.  The mock observational catalogues were produced
assuming that all clusters with $M_{200}$ above $4.6\times10^{14}h^{-1}\Msun$
are detected, and none below.  In all they have on average 41 clusters (a number
similar to the {\it HIFLUGCS} sub-sample we are working with). 

\begin{figure}
\centering
\psfig{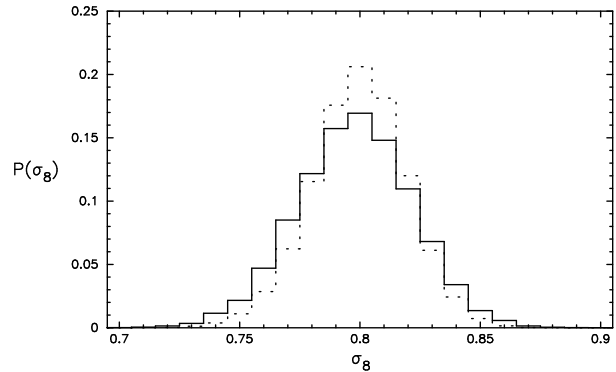}
\caption[sig8]{Marginalized probability distributions for $\sigma_{8}$, obtained 
through the 2D Kolmogorov--Smirnov test (full line) and the likelihood method 
(dotted line).}
\end{figure}

When applying the 2D Kolmogorov--Smirnov
test to each of the 1000 mock observational catalogues, 1000 synthetic
catalogues were produced for each $\sigma_{8}$ in the interval of interest
[0.60,1.00], thus overall around $4\times10^{7}$ catalogue comparisons were
made. In this case, the likelihood function is the product of the Poisson 
probabilities of finding exactly one cluster 
in the element $dM_{200}dz$ at each of the $(M_{200}^i,z^i)$ combinations 
present in the mock catalogues, 
and of finding zero clusters elsewhere in the $(M_{200},z)$ plane 
(see e.g Marshall et al.~1983).

In Fig.~1 we show the probability distributions for $\sigma_{8}$ obtained by the 
two methods. This comparison shows that both methods are unbiased, picking up 
the 
fiducial
$\sigma_{8}=0.8$ as the most probable value.  Further, the shape of the two
probability distributions is very similar, though applying the 2D
Kolmogorov--Smirnov test seems to result in slightly more conservative
confidence limits.  We have made simulations with other initial assumptions and
the results do not change qualitatively.

\begin{figure}
\centering
\psfig{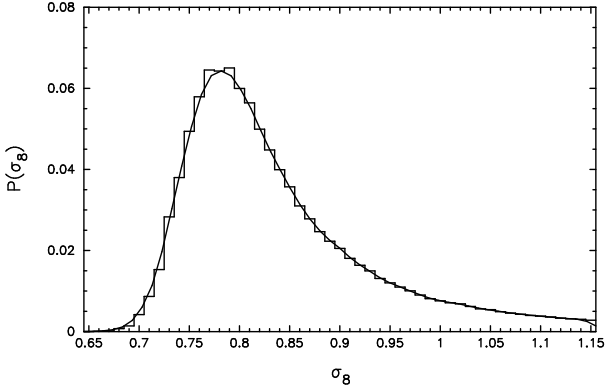}
\caption[sig8]{Marginalized probability distribution for $\sigma_{8}$, showing 
the bins as calculated and a smoothed version of the distribution.}
\end{figure}

\subsection{Application to {\it HIFLUGCS}}

The application of the 2D Kolmogorov--Smirnov test to the {\it HIFLUGCS} 
data in the manner previously described results 
in the marginalized probability distributions,  
for each free parameter over the three others, presented in Figs.~2 and 3. 
The histograms originate from the discretization of our parameter space 
($\sigma_{8},\,A,\,\alpha,\,\sigma_{\log_{10}L_{\rm X}}$) for the Monte 
Carlo simulations. The continuous lines represent the most probable underlying 
probability density functions, and result from the application of a 
non-parametric smoothing technique to the histogram data. Note that these 
functions have been renormalized for easier comparison with the histograms. 
In summary: 
\begin{equation}
\sigma_{8}\simeq0.78\,\,{\rm within}\,\,[0.72,1.08]\,,
\end{equation}
and
\begin{equation}
\log_{10}(L_{\rm X}/h^{-2}\;{\rm erg}\,{\rm s}^{-1})=A+\alpha\log_{10}(kT/{\rm 
keV})\,,
\end{equation}
with
\begin{equation}
A\simeq42.1 ,\,{\rm within}\,\,[41.2,42.5]  \,,
\end{equation}
\begin{equation}
\alpha\simeq2.5\,\,{\rm within}\,\,[1.5,3.5]\,,
\end{equation}
\begin{equation}
\sigma_{\log_{10}L_{\rm X}}\simeq0.3\,\,{\rm within}\,\,[0.0,0.6]\,.
\end{equation}
where the given ranges are all at the 95 per cent confidence
level. The most probable combination of the four parameters we
consider is $\sigma_8=0.77$, $A=42.2$, $\alpha=2.6$, and
$\sigma_{\log_{10}L_{\rm X}}=0.175$.  Note that the distribution for
$\sigma_8$ is considerably non-gaussian, with the median value
$\sigma_8=0.81$ being higher than the modal one. The tail
extends much further to high $\sigma_8$ because, as the number of
existing clusters increases, it remains possible to reproduce the
observed number of clusters by simultaneously choosing lower values
for $A$ and higher values for $\sigma_{\log_{10}L_{\rm X}}$. In the
limit where no dispersion in the relation between X-ray luminosity and
temperature is allowed, the possibility of $\sigma_8$ taking high
values disappears, and the marginalized probability distribution for
$\sigma_8$ becomes close to gaussian. Re-doing our analysis not
allowing for any dispersion in the relation between X-ray luminosity
and temperature, the most probable value for $\sigma_8$ changes to
0.76, with the 95 per cent confidence interval now extending from 0.70
to 0.81, while the most probable values for the parameters $A$ and
$\alpha$ stay almost the same, changing to respectively $42.3$ and
$2.6$.

\begin{figure}
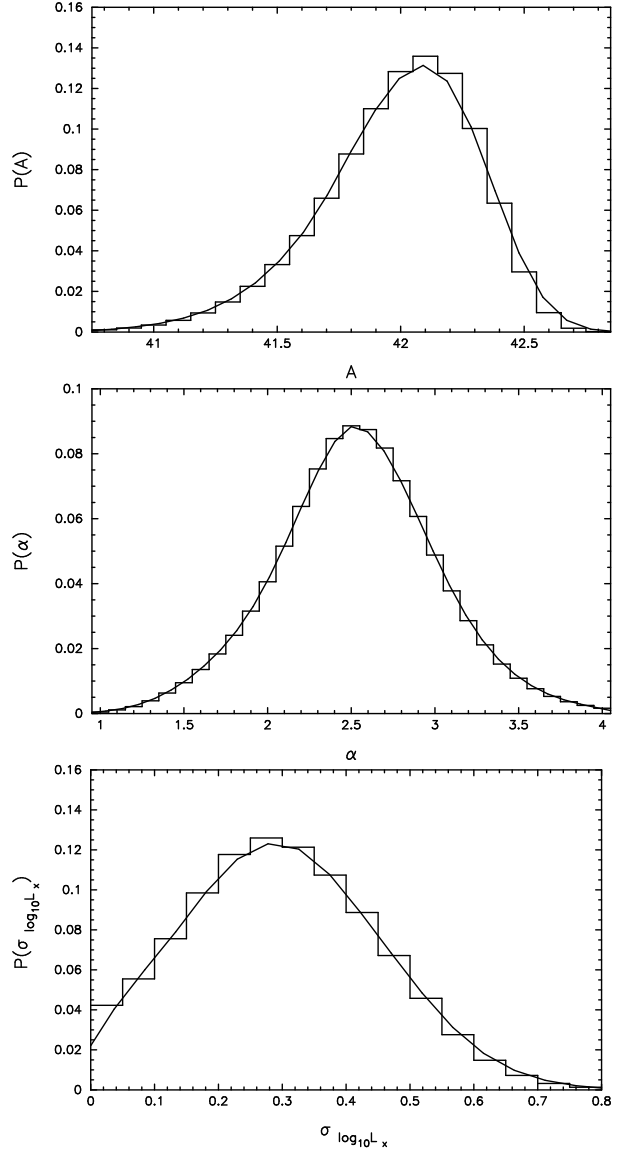

\centering
\psfig{figure=Fig3.ps,width=8cm}
\psfig{figure=Fig4.ps,width=8cm}
\psfig{figure=Fig5.ps,width=8cm}
\caption[LTrelation]{Marginalized probability distributions for 
the normalization $A$ (top) and slope $\alpha$ (middle) of the 
relation between X-ray temperature and luminosity, as well as for 
its dispersion $\sigma_{\log_{10}L_{\rm X}}$ (bottom).}
\end{figure}

These results are similar to those obtained by Ikebe et al.~(2002).
The comparison between the two analysis is made difficult by the fact
that they only indicate the most probable values for $A$, $\alpha$,
and $\sigma_{\log_{10}L_{\rm X}}$ for their best-fit $\Omega_0$ and
$\sigma_8$, which are $0.26$ and $0.94$ respectively if only $T>3$ keV
clusters are considered.  Concentrating on this case, and assuming
$\Omega_0=0.26$ (plus $\Gamma=0.206$, as in Ikebe et al.~2002), we
attempted to recover the values obtained by Ikebe et al.~(2002) for
the other 4 parameters.  We took into account that their assumed
normalization for the cluster mass ($M_{200}$) to X-ray temperature
relation (estimated as if $z=0.05$) is somewhat higher, such that for
a 3 keV cluster they assume a cluster mass around 4 per cent higher
than we, while at 6 keV the difference increases to 13 per cent, as
well as the fact that they do not take into account a possible
dispersion in the cluster X-ray temperature at fixed mass.  A most
probable value of $\sigma_8=0.98$ was obtained by applying our
procedure, with good agreement also found for the parameters $A$,
$\alpha$, and $\sigma_{\log_{10}L_{\rm X}}$.  Given that some
differences remain between the two analysis, our results thus seem to
be consistent with those of Ikebe et al.~(2002).

In order to determine which type of information in the data is driving
the results, we determined the most probable values for the four
parameters under consideration by applying in isolation the 2D
Kolmogorov--Smirnov test to the three available distributions of
cluster properties --- ($z,kT$), ($z,L_{\rm X}$) and ($kT,L_{\rm X}$).
$\sigma_8$ is essentially unconstrained by the ($z,L_{\rm X}$)
distribution.  All the information comes from the two others, with the
($kT,L_{\rm X}$) distribution being slightly more constraining than
the ($z,kT$) one.  Consistently, the former prefers $0.78$ as the most
probable value for $\sigma_8$, while the latter settles for $0.79$.
The information on the parameters $A$, $\alpha$, and
$\sigma_{\log_{10}L_{\rm X}}$ is roughly equally distributed amongst
the three distributions, though again ($kT,L_{\rm X}$) and ($z,L_{\rm
X}$) are always the most and least constraining respectively, and when
taken in isolation all three distributions lead to very similar
results.

In Fig. 4 we compare the cluster properties between a realization of
the {\it HIFLUGCS} sub-sample selected for the analysis, the mock
sample that most resembles it, generated for the most probable set of
parameters, and the underlying cluster population. Notice that the
incompleteness of the flux-limited samples increases considerably as
the cluster X-ray temperature gets lower, so that below a X-ray
temperature of about 5 keV we can conclude that {\it HIFLUGCS} is
vastly incomplete.

\begin{figure}
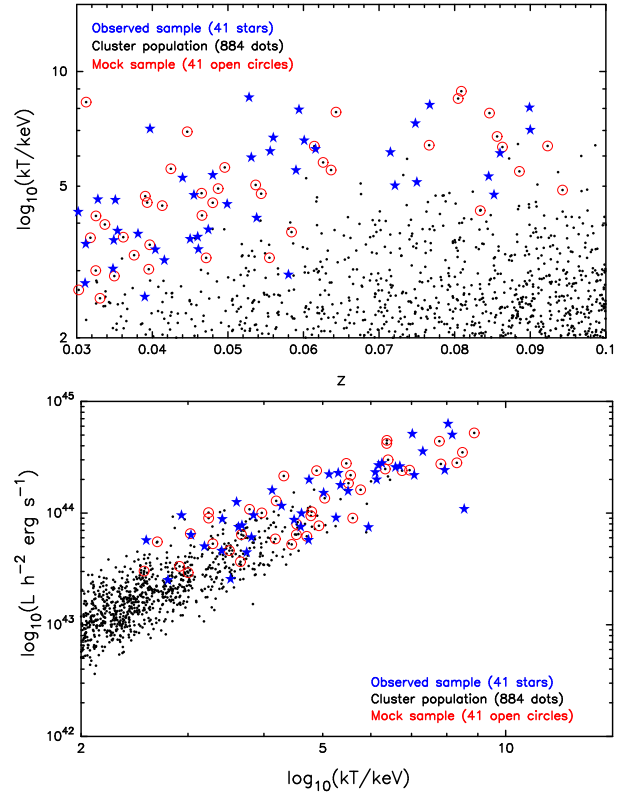

\centering
\psfig{figure=Fig6.ps,width=8cm}
\psfig{figure=Fig7.ps,width=8cm}
\caption[Clusterproperties]{Comparison of distribution of cluster 
properties ($T$ vs. $z$, top; $L_{\rm X}$ vs. $T$, bottom) between 
a realization of the {\it HIFLUGCS} sub-sample considered, the mock sample 
that most resembles it, generated for the most probable set of 
parameters, and the underlying cluster population.}
\end{figure}

Our analysis is for $\Omega_0 = 0.35$, being the value for which the large 
hydrodynamical simulations were run. The spectacular recent results from WMAP 
(Bennett et al.~2003; Spergel et al.~2003) are consistent with this, but their 
best-fit is somewhat lower at $\Omega_0 = 0.27$. While we are unable to run new 
large simulations, we can predict the effect on $\sigma_8$ using the scaling 
found in earlier analyses. VL99 found that for flat cosmologies $\sigma_8 
\propto \Omega_0^{-0.47}$, and using this scaling we obtain a best-fit 
$\sigma_8$ of 0.88 for $\Omega_0 = 0.27$. Given the small range over which this 
scaling is needed, the fractional uncertainty in $\sigma_8$ should be unchanged.

\section{Discussion}

To set the context for the following discussion, we first remind the
reader of the constraint from VL99, which for
$\Omega_0 = 0.35 $ gave $\sigma_8 = 0.92$ within [0.73, 1.12] 
at 95 per cent confidence. By contrast, in Seljak (2002) a value of 
$\sigma_8 =0.70$ was obtained, based on the cluster mass to X-ray 
temperature relation derived in Finoguenov et al.~(2001) from cluster data. 

The calculation of $\sigma_8$ performed in this paper is 
substantially different from that carried out in VL99:  
the semi-analytical modelling featured a change in the normalization of the 
assumed cluster mass--temperature relation and in the shape of the 
assumed cluster mass function; a different observational dataset was used; 
and the method of estimating $\sigma_8$ from Monte Carlo simulations differed 
from the previous likelihood-type calculation, where no dispersion in the 
cluster relations was considered. Although, the most probable value for 
$\sigma_8$ is quite different in both cases, the 
95 per cent confidence intervals happen to be very similar. 

In order to find the most important factors behind the different results, 
we ran several Monte Carlo simulations. Firstly, we found that the inclusion of
dispersion in the mass--temperature relation at the level considered in this 
paper does not seem to make much difference. Secondly, and more surprisingly, 
we found that replacing the Evr02 mass function with the Press--Schechter or 
the Jenkins et al.~(2002) mass function also changes the most probable value for 
$\sigma_8$ by less than two per cent. This appears in 
contradiction with claims in the literature, including our own 
(Wu 2001; Pierpaoli et al.~2001; Viana et al.~2002), 
that the choice of mass function can change $\sigma_8$ by five to ten 
percent.  However that statement is only true if one keeps the 
mass--temperature relationship unchanged, but in fact these
different mass functions refer to different masses; Evr02 gives the number
density of haloes with mass $M_{200}$, Press--Schechter uses the virial mass
which for the cosmology assumed here is about $M_{108}$, and the
Jen01 mass function corresponds to $M_{63}$ for the same cosmology.  If we
use the NFW cluster density profile to scale these mass functions to the same
mass definition (e.g.~the virial mass), most of the difference in $\sigma_8$
disappears.  We note however that this similarity of results may be specific to
the cosmology adopted here.

A change that does make a difference is that, as compared to VL99, this paper 
uses a much lower normalization of the cluster mass to X-ray temperature 
relation. So that we could determine the influence of such normalization 
on our results, and be able to compare them more easily with others, we 
calculated the dependence of the most probable $\sigma_8$ on the value of 
the assumed present-day mass ($M_{200}@5{\rm keV}$) of a 5 keV cluster 
(approximately the median 
temperature of the {\it HIFLUGCS} sub-sample we work with). Taking the index of 
the 
cluster mass to X-ray temperature relation to be the standard 1.5, we found 
\begin{equation}
\sigma_8=0.37+0.11 \times \left(\frac{M_{200}@5{\rm keV}}{10^{14}h^{-1}\Msun} 
\right)^{0.83} \,.
\end{equation} 
Note that in our main calculation we assumed $M_{200}@5{\rm
keV}=4.83\times10^{14}h^{-1}\Msun$, for an index of 1.54.  Given that in VL99 it
was assumed $M_{200}@5{\rm keV}=7.67\times10^{14}h^{-1}\Msun$, we
obtain $\sigma_8=0.97$ as the value we would expect from VL99 if the only
significant difference between the analyses was that change in the
normalization.  Comparing with the VL99 value of $\sigma_8=0.92$, this seems to
be correct, with the {\it HIFLUGCS} sub-sample we consider favouring just a
slightly higher normalization than the Henry \& Arnaud (1991) dataset used in
VL99.  Although it is difficult to untangle all the competing effects, we
suspect that together the new analysis method and the {\it HIFLUGCS} dataset
allow for a much better estimate of incompleteness which would help explain why
they favour a higher normalization.

Turning to comparison with other work, the reason why Seljak (2002) obtained a
significantly smaller value for $\sigma_8$ with relation to VL99 (to which it is
more easily compared), is the assumption at fixed cluster temperature of a mass
that is about 2.4 times lower than that assumed in VL99, though this effect is
mitigated by Seljak's assumed local cluster abundance at about 6 keV (from
Pierpaoli et al.~2001) which was higher than that of VL99.  In this paper we too
have a cluster mass at 6 keV which is much smaller than VL99, but the reduction
is by a smaller factor of 1.6.

As we were completing this work, a paper by Pierpaoli et al.~(2003)
appeared in which a similar analysis to ours and that in Ikebe et
al.~(2002) was carried out. The observed cluster sample is also
derived from {\it HIFLUGCS}, but otherwise they use a different
approach to obtain constraints on $\sigma_8$. While both here and in
Ikebe et al.~(2002) it is attempted to constrain $\sigma_8$
simultaneously with the X-ray temperature to luminosity relation, in
Pierpaoli et al.~(2003) such a relation is assumed {\em a priori} (to
be that given by expression~3 in Ikebe et al.~2002). We have attempted
to reproduce the constraint obtained for $\sigma_8$ by Pierpaoli et
al.~(2003) when they derive the observed cluster sample just from {\it
HIFLUGCS}.  Such a constraint can be read from the full line in
Figs.~4 and 5 of Pierpaoli et al.~(2003). Concentrating on the case of
$\Omega_0=0.35$, and performing an analysis equivalent to that in
Pierpaoli et al.~(2003), taking care to make the same assumptions and
apply the selection criteria in the same manner, but using the 2D-KS
method instead, we found a most probable value for $\sigma_8$ and the
90 per cent confidence interval very similar to theirs (a slight
overestimation by $0.02$). On the other hand, if we just change in our
own analysis the cluster mass to X-ray temperature relation so that
$M_{200}@5{\rm keV}=3.82\times10^{14}h^{-1}\Msun$ at present and
its index to 1.5, as in Pierpaoli et al.~(2003), the result for
$\sigma_8$ is again very similar (about $0.71$) to that obtained in
Pierpaoli et al.~(2003). Clearly, the most significant factor leading
to the different {\it HIFLUGCS} based result obtained here and in
Pierpaoli et al.~(2003) regarding $\sigma_8$, is the difference in the
normalization of the cluster mass to X-ray temperature relation.

Our results do not indicate a dramatic reduction in $\sigma_8$ derived from 
the abundance of X-ray clusters. Several other recent analyses have favoured 
low $\sigma_8$, for instance from 2dF and CMB data (Efstathiou et al.~2002; 
Lahav et al.~2002), from using weak lensing to estimate cluster masses 
(Viana et al.~2002), and from the local X-ray cluster luminosity function (Allen 
et al.~2002), but those are at least marginally compatible 
with our present result given the uncertainties. Indeed, results from WMAP have 
forced a modest increase in estimates of $\sigma_8$ via CMB data (Spergel et 
al.~2003).
Our estimated value for $\sigma_8$ is compatible with all published weak 
lensing measurements (e.g Bacon et al.~2002; H\"{o}kstra et al.~2002; 
Refregier, Rhodes \& Groth 2002; Van Waerbeke et al.~2002), though only 
marginally 
with the very low results of Brown et al.~(2003), Hamana et al.~(2002) 
and Jarvis et al.~(2003), as well as at the other extreme with that of 
Maoli et al.~(2001). 

In the near future, a decrease in the uncertainty in the estimation of 
$\sigma_8$ from X-ray clusters could come from essentially two sources. On the 
theoretical side, 
it would be important to reliably estimate the X-ray luminosity of clusters 
using hydrodynamical $N$-body simulations. This would enable one to bypass 
the X-ray temperature as the cluster mass estimator. Though temperature is more 
reliable, it ends up not being as useful as it could be due to the fact 
that all cluster catalogues are flux-limited instead of temperature 
selected, so an estimation of the cluster X-ray flux always needs to be 
made. On the observational side, both an improvement in the temperature 
determination and an increase in the range of redshift probed 
(i.e. a decrease in the X-ray flux detection limit) would help bring 
down the uncertainty in the estimation of $\sigma_8$. Hopefully, both 
can be achieved with the X-ray satellites {\em Chandra} and 
{\em XMM-Newton}. In particular, it is expected that the serendipitous 
cluster survey {\em XCS} (Romer et al.~2001) to be assembled with {\em 
XMM-Newton} data will help greatly in both issues.

\section*{Acknowledgements}

The simulations used in this paper were carried out on the Cray-T3E
at the EPCC as part of the Virgo Consortium programme of
investigations into the formation of structure in the Universe. 
STK is supported by PPARC and ARL in part by the Leverhulme Trust. 
During the initial preparation of this paper, OM was supported by a DPST
scholarship from the Thai government and PAT was a PPARC Lecturer
Fellow. PTPV acknowledges the financial support of FCT through 
project POCTI/FNU/43753/2001 (partially funded through FEDER). 
We thank Elena Pierpaoli for useful discussions.

\end{document}